\documentclass[12pt]{article}
\usepackage{epsf}
\usepackage{amsmath,amssymb}
\usepackage{slashed}

\usepackage{graphicx}

\allowdisplaybreaks

\usepackage{comment}

\begin{document}

\newcommand{\lsim}{\stackrel{<}{_\sim}}
\newcommand{\gsim}{\stackrel{>}{_\sim}}

\newcommand{\rem}[1]{{$\spadesuit$\bf #1$\spadesuit$}}

\renewcommand{\thefootnote}{\fnsymbol{footnote}}
\setcounter{footnote}{0}

\begin{titlepage}

\def\thefootnote{\fnsymbol{footnote}}

\begin{center}

\hfill UT-15-40\\
\hfill KEK-TH-1875\\

\vskip .75in

{\large \bf 

  Renormalization-Scale Uncertainty
  \\
  in the Decay Rate of False Vacuum

}

\vskip .25in

{\large
  Motoi Endo$^{\rm (a,b)}$, Takeo Moroi$^{\rm (a,b)}$, 
  Mihoko M. Nojiri$^{\rm (b,c,d)}$,\\
  and Yutaro Shoji$^{\rm (a)}$
}

\vskip 0.25in

$^{\rm (a)}${\em 
Department of Physics, University of Tokyo, Tokyo 113-0033, Japan}

\vskip 0.1in

$^{\rm (b)}${\em 
Kavli IPMU (WPI), University of Tokyo, Kashiwa, Chiba 277-8583, Japan}

\vskip 0.1in

$^{\rm (c)}${\em 
KEK Theory Center, IPNS, KEK, Tsukuba, Ibaraki 305-0801, Japan}

\vskip 0.1in

$^{\rm (d)}${\em 
SOKENDAI (The Graduate University for Advanced Studies),\\
Tsukuba, Ibaraki 305-0801, Japan}

\end{center}
\vskip .5in

\begin{abstract}

  We study radiative corrections to the decay rate of false vacua,
  paying particular attention to the renormalization-scale dependence
  of the decay rate.  The decay rate exponentially 
  depends on the bounce action.  The bounce action itself is
  renormalization-scale dependent. To make the decay rate
  scale-independent, radiative corrections, which are due to the
  field fluctuations around the bounce, have to be included.  We show
  quantitatively that the inclusion of the fluctuations suppresses the
  scale dependence, and hence is important for the precise calculation of
  the decay rate.  We also apply our analysis to a supersymmetric
  model and show that the radiative corrections are important for
  the Higgs-stau system with charge breaking minima.

\end{abstract}

\end{titlepage}

\renewcommand{\thepage}{\arabic{page}}
\setcounter{page}{1}
\renewcommand{\thefootnote}{\#\arabic{footnote}}
\setcounter{footnote}{0}
\renewcommand{\theequation}{\thesection.\arabic{equation}}

\section{Introduction}
\label{sec:intro}
\setcounter{equation}{0}

In particle physics and cosmology, decay of false vacua is an
important subject.  For example, with the observed Higgs and top
masses, it has been known that the Higgs quartic coupling constant
becomes negative above $\sim 10^{10}$\,GeV if the standard
model (SM) is a good effective theory up to
the scale \cite{Degrassi:2012ry}. Then the electroweak symmetry breaking (EWSB) vacuum is a
false vacuum. Even if there exists a true vacuum other than the EWSB vacuum, we may
still live in the EWSB vacuum as long as the lifetime
of the EWSB vacuum is longer than the present cosmic time.  In models beyond the SM, the EWSB
vacuum may still be a false vacuum.  For example, in supersymmetric
(SUSY) models, there may exist a color and/or charge breaking (CCB)
vacuum (at which some of the superpartners of quarks and/or leptons
acquire non-vanishing expectation values) whose vacuum energy is
lower than that of the EWSB vacuum.  Existence of such a CCB vacuum
imposes important and stringent bounds on SUSY models \cite{Frere:1983ag,Gunion:1987qv,Casas:1995pd,Kusenko:1996jn}.

Precise calculation of the decay rate of the false vacua is important from both theoretical and phenomenological points
of view.  The procedure to calculate the decay rate 
was formulated in \cite{Coleman:1977py, Callan:1977pt}, in which the
decay rate is evaluated by performing the path integral around the
saddle-point solution (i.e., so-called the ``bounce'') of the equation
of motion in the Euclidean field theory.  Given the bounce solution, 
the decay rate per unit volume is given by
\begin{align}
  \gamma \equiv 
  {\cal A} e^{-{\cal B}},
  \label{decayrate}
\end{align}
where ${\cal B}$ is the bounce action, which is the Euclidean action
of the bounce solution, while the prefactor ${\cal A}$ takes account
of the effects of fluctuations around the bounce.  In many analyses,
${\cal B}$ has been evaluated from the tree-level Lagrangian, while an
order-of-magnitude estimate has been adopted for ${\cal A}$.  The main
subject of this paper is the calculation of ${\cal A}$, which is
important to determine the overall scale of the decay rate.  Another
motivation of the calculation comes from the scale independence of the
decay rate.  ${\cal B}$ inevitably depends on the renormalization
scale $Q$ at which the tree-level parameters in the Lagrangian are
defined.  As we will see, the scale dependence of ${\cal B}$ can be
sizable.  The decay rate of the false vacuum is physical quantity, and
therefore, the scale dependence should be cancelled in the expression
of $\gamma={\cal A} e^{-{\cal B}}$.

In this paper, we discuss the calculation of the decay rate of false
vacua, paying particular attention to the renormalization-scale
dependence of the decay rate $\gamma$.  In Section
\ref{sec:formalism}, we summarize the formalism to calculate the
prefactor.  In Sections \ref{sec:scalar} and
\ref{sec:mssm}, we perform numerical calculations of the
decay rate $\gamma$ for a simple model of a real scalar field 
and Higgs-stau system in the minimal SUSY SM (MSSM), respectively.  We show that, in those
models, ${\cal B}$ has sizable dependence on $Q$, while the scale
dependence of $\gamma={\cal A}e^{-{\cal B}}$ becomes weak once
the effect of the prefactor ${\cal A}$ is properly taken into account.
Section \ref{sec:summary} is devoted for the summary of this paper.

\section{Formalism}
\label{sec:formalism}
\setcounter{equation}{0}

In order to calculate the decay rate of false vacua, we follow
the procedure given in \cite{Coleman:1977py,Callan:1977pt,Avan:1985eg,Isidori:2001bm,Baacke:2003uw,Dunne:2007rt,Dunne:2005rt}.  
In the calculation, the bounce solution plays an important role. The
bounce is the solution of the classical field equations that
interpolates between the false and true vacua.  It is an
$O(4)$ symmetric solution of the four-dimensional Euclidean equation of
motion, and it only depends on the radial distance in the
Euclidean space $r=\sqrt{x_\mu x_\mu}$.  In the following, the bounce
solution is denoted as $\sigma(r)$ (or $\sigma_i$, when we need to
specify the individual fields). We also denote the expectation value 
of the scalar field at the false vacuum as $\bar{\sigma} \equiv \sigma(r \to \infty)$.

Hereafter, we calculate the prefactor ${\cal A}$ at the one-loop level.
We consider the prefactor arising from the coupling
of the bounce to scalar and spinor fluctuations. Then, the prefactor ${\cal A}$ can be decomposed as
\begin{align}
  {\cal A} = \frac{{\cal B}^2}{4\pi^2}
  A'_\phi A_\psi,
  \label{PrefactorA}
\end{align}
where $A'_\phi$ and $A_\psi$ are scalar- and fermion-loop
contributions, respectively.  As we see below, $A_\psi$ is
dimensionless, while the mass dimension of $A'_\phi$ is four.

We assume the bosonic contribution arises from the Euclidean Lagrangian of
the following form:
\begin{align}
  {\cal L}_{\phi} = \frac{1}{2} \partial_\mu \phi_i \partial_\mu \phi_i
  + V(\sigma, \phi_i),
\end{align}
where $\phi_i$ denotes scalar fluctuations around the bounce 
solution $\sigma(r)$, and $V$ is the scalar potential.  We take the
basis of the scalar fields such that each $\phi_i$ becomes a mass
eigenstate around the false vacuum.  Then,
\begin{align}
  A'_\phi =
  \left| 
    \frac{ {\rm Det}' [-\partial^2 + V_{ij}(\sigma)]}
    {{\rm Det} [-\partial^2 + \bar{V}_{ij}]}
  \right|^{-1/2}
  e^{-S_\phi^{\rm (c.t.)}},
  \label{A_phi}
\end{align}
where
\begin{align}
  V_{ij}(\sigma) \equiv
  \left.
    \frac{\partial^2 V}{\partial \phi_i \partial \phi_j}
  \right|_{\phi=0},
\end{align}
$\bar{V}_{ij} \equiv V_{ij}(\bar{\sigma})$, 
and $S_\phi^{\rm (c.t.)}$ is the counter term to remove the
divergences due to $\phi_i$.  In addition, ${\rm Det}'$ is 
the functional determinant with omitting four zero-eigenvalues
associated with the translation of the bounce solution.  
Then, the mass dimension of $A'_\phi$ is four, that is the
mass dimension of $\gamma$.
$A'_\phi$ is often estimated to be the fourth power of a typical
mass scale in the bounce.  

The fermionic part of the Euclidean Lagrangian is denoted as
\begin{align}
  {\cal L}_{\psi} = \bar{\psi} \gamma_\mu \partial_\mu \psi +
  M (\sigma) \bar{\psi} \psi,
\end{align}
where $\gamma_\mu$ is the $\gamma$-matrix, satisfying the
anti-commutation relation as
$\{\gamma_\mu,\gamma_\nu\}=2\delta_{\mu\nu}$.  Then, fermionic 
contribution is given by
\begin{align}
  A_\psi =
  \left[
    \frac{ {\rm Det} [-(\slashed{\partial}+M)(\slashed{\partial}-M)]}
    {{\rm Det} [-\partial^2 + \bar{M}^2]}
  \right]^{1/2}
  e^{-S_\psi^{\rm (c.t.)}},
  \label{A_psi}
\end{align}
where $\bar{M}\equiv M(\bar{\sigma})$, and $S_\psi^{\rm (c.t.)}$ is the counter term.

We first discuss the effect of fluctuations which are not related to
the zero-modes.  As shown in Eqs.~\eqref{A_phi} and \eqref{A_psi}, the
prefactor ${\cal A}$ is related to the following quantity:
\begin{align}
  A_\varphi = 
  \left(
    {\rm Det}
    \left[ 
      \frac{ -\partial^2 + W (r) }
      { -\partial^2 + \bar{W} }
    \right]
  \right)^{(-1)^{F+1}/2}
  e^{-S_\varphi^{\rm (c.t.)}},
\end{align}
with $\varphi = \phi$ or $\psi$,
where $(-1)^{F}=+1$ and $-1$ for boson and fermion, respectively, 
$S_\varphi^{\rm (c.t.)}$ is the counter term, and $\bar{W} \equiv W(r \to \infty)$, 
which is the value at the false vacuum. 

We can obtain a formal expression of $A_\varphi$.  Expanding $W$ as
\begin{align}
  W (r) = \bar{W} + \delta W (r),
\end{align}
we obtain
\begin{align}
  \ln A_\varphi
  = - \sum_{p=1}^\infty s_\varphi^{(p)},
\end{align}
where
\begin{align}
  s_\varphi^{(p)} \equiv \frac{(-1)^{F+p+1}}{2p} {\rm Tr}
  \left[ 
    \delta W \frac{1}{-\partial^2 + \bar{W}}
  \right]^p
  + (\mbox{counter term}),
\end{align}
with ``Tr'' denoting the functional trace.  In addition, divergences
appear only for $p=1$ and $2$, and hence the counter term contributions 
do not appear for $p\geq 3$.  In our analysis, $s_\varphi^{(1)}$ and
$s_\varphi^{(2)}$ are evaluated by performing the momentum 
integration with the $\overline{\rm MS}$ scheme:
\begin{align}
  s_\varphi^{(1)} &= (-1)^{F+1}\sum_i\delta \tilde{W}_{ii}(0)
  \frac{\bar{W}_{ii}}{32\pi^2}\left[1-\ln\frac{\bar{W}_{ii}}{Q^2}\right],
  \label{s1}
  \\
  s_\varphi^{(2)} &= (-1)^{F+1}\frac{1}{512\pi^4}
  \sum_{i,j}
  \int\! dk\, k^3\delta\tilde{W}_{ij}(k)\delta \tilde{W}_{ji}(k)\nonumber\\
  &\hspace{3ex}\times
  \left[
    2
    -\frac{1}{2}\ln\frac{\bar{W}_{ii}\bar{W}_{jj}}{Q^4}
    +\frac{\bar{W}_{ii}-\bar{W}_{jj}}{2k^2}\ln\frac{\bar{W}_{ii}}{\bar{W}_{jj}}
    -\frac{\omega^2}{2k^2}\ln
    \frac{k^2+\bar{W}_{ii}+\bar{W}_{jj}+\omega^2}
    {k^2+\bar{W}_{ii}+\bar{W}_{jj}-\omega^2}
  \right],
  \label{s2}
\end{align}
where
\begin{align}
 \omega^2&=\sqrt{(\bar{W}_{ii}+\bar{W}_{jj}+k^2)^2-4\bar{W}_{ii}\bar{W}_{jj}},
 \\
 \delta \tilde{W}(k)&=\frac{4\pi^2}{k}\int\! dr\,r^2\delta W(r)J_1(kr),
\end{align}
with $J_1(x)$ being the modified Bessel function of the first kind.
Notice that $\bar{W}$ is a diagonal matrix in our choice of the basis.

Next, let us consider the finite part, i.e., $\sum_{p\geq
  3}s_\varphi^{(p)}$.  Because the bounce solution has O(4)
symmetry, the eigenfunctions of the operator $(-\partial^2+ W)$ can be
characterized by the quantum numbers for the rotational group of the
four-dimensional Euclidean space, i.e., SU(2)$_A\times$ SU(2)$_B$.  We
denote the spin operators for SU(2)$_A$ and SU(2)$_B$ as $\hat{A}_i$
and $\hat{B}_i$, respectively, and the eigenvalues of
$(\hat{A}^2,\hat{A}_3,\hat{B}^2,\hat{B}_3)$ are denoted as
$(j_A,m_A,j_B,m_B)$; $j_A=j_B$ for scalars, and
$j_A=j_B\pm\frac{1}{2}$ for fermions.  Hereafter, we denote
\begin{align}
  J \equiv {\rm min}(j_A,j_B),
\end{align}
which takes the values of $J=0$, $\frac{1}{2}$, $1$, $\frac{3}{2}$,
$\cdots$.  Then, the functional determinant of our interest can be
decomposed into the contributions of each $J$ as
\begin{align}
  {\rm Det} 
  \left[
    \frac{ -\partial^2 + W }
    { -\partial^2 + \bar{W} }
  \right]
  = 
  \prod_J 
  {\rm Det} 
  \left[
    \frac{ - \Delta_J + W }
    { - \Delta_J + \bar{W} }
  \right],
\end{align}
where $\Delta_J$ is the four-dimensional Laplace operator acting on
the mode with $J={\rm min}(j_A,j_B)$.  For scalars,
\begin{align}
  [\Delta_J - W ]_\phi = 
  \partial_r^2 + \frac{3}{r} \partial_r 
  - \frac{2J(2J+2)}{r^2} 
  - V_{ij},
\end{align}
and for fermions,
\begin{align}
  [\Delta_J - W ]_\psi = 
  \partial_r^2 + \frac{3}{r} \partial_r 
  - 
  \left( 
    \begin{array}{cc}
      2J (2J + 2) r^{-2} + M^2 & \partial_r M \\
      \partial_r M & (2J+1) (2J+3) r^{-2} + M^2
    \end{array}
  \right).
\end{align}

Using the technique given in \cite{Vleck,Cameron,Gelfand,Dashen:1974ci,Coleman:1978ae,Kirsten:2003py,Kirsten:2004qv}, it is possible to express the determinant as follows,\footnote
{Here and hereafter, $(\varphi_J / \bar{\varphi}_J)$ should be
  understood as the product $\varphi_J\bar{\varphi}_J^{-1}$ if
  $\varphi_J$ and $\bar{\varphi}_J$ are matrices.}
\begin{align}
  {\rm Det} 
  \left[
    \frac{ - \Delta_J + W }
    { - \Delta_J + \bar{W} }
  \right]
  = 
  \left.
    {\rm det} 
    ( \varphi_J / \bar{\varphi}_J )^{N_J}
  \right|_{r=\infty},
  \label{Det_J}
\end{align}
where $N_J$ is the degeneracy; $N_J=(2J+1)^2$ for a scalar, and
$N_J=2(2J+1)(2J+2)$ for a fermion.  Notice that the factor of 2 in 
$N_J$ for fermions originates from two choices of $j_A=j_B-\frac{1}{2}$
and $j_A=j_B+\frac{1}{2}$.
In addition, $\varphi_J$ is the
function, which is regular in $r=0$, obeying the following equation:
\begin{align}
  [\Delta_J - W(r) ] \varphi_J (r) = 0.
  \label{EoM_u}
\end{align}
The function $\bar{\varphi}_J$, which has the same boundary condition
as $\varphi_J$ at $r=0$, is obtained from Eq.~\eqref{EoM_u} with
$W$ being replaced by $\bar{W}$.  We define the function
$\varphi_J^{(p)}$ which obeys
\begin{align}
  [ \Delta_J - \bar{W} ] \varphi_J^{(p)} =
  \delta W \varphi_J^{(p-1)},
  ~~~ (p\geq 1),
\end{align}
with $\varphi_J^{(0)}=\bar{\varphi}_J$.  Then, $\varphi_J =
\sum_{p=0}^\infty \varphi_J^{(p)}$, and the following relation holds:
\begin{align}
  \sum_{p\geq 3} s_\varphi^{(p)} = 
  \frac{(-1)^F}{2}\sum_J N_J
  \left.
    \left[
      {\rm tr} \ln ( \varphi_J / \bar{\varphi}_J )
      - \tilde{\varphi}_J
    \right] 
  \right|_{r=\infty},
  \label{s3-infty}
\end{align}
where
\begin{align}
  \tilde{\varphi}_J \equiv
  {\rm tr} \left[
    (\varphi_{J}^{(1)} / \bar{\varphi}_J)
    - \frac{1}{2} (\varphi_{J}^{(1)} / \bar{\varphi}_J)^2
    + (\varphi_{J}^{(2)} / \bar{\varphi}_J)
  \right].
\end{align}

Using Eqs.~\eqref{s1}, \eqref{s2}, and \eqref{s3-infty},
$A_\varphi$ is given by
\begin{align}
  A_\varphi = 
  e^{-s_\varphi^{(1)}-s_\varphi^{(2)}}
  \prod_J
  \left.
    \left[
      {\rm det} ( \varphi_J / \bar{\varphi}_J )
      e^{- \tilde{\varphi}_J}
    \right]^{(-1)^{F+1}N_J/2}
  \right|_{r=\infty}.
  \label{A_varphi}
\end{align}
This expression can be used for numerical calculations.  Importantly,
the quantities $s_\varphi^{(1)}$ and $s_\varphi^{(2)}$ are
finite, while the quantity ${\rm det}(\varphi_J/\bar{\varphi}_J)e^{- \tilde{\varphi}_J}$
approaches to $1$ as $J\rightarrow\infty$, which make it possible to
numerically evaluate $A_\varphi$ with Eq.~\eqref{A_varphi}.

In general, the bounce action cannot be expressed by
analytic functions. For our numerical calculations in the following sections, 
we use {\tt CosmoTransitions}~2.01a \cite{Wainwright:2011kj} to determine the bounce
solution as well as ${\cal B}$. 

In the calculation of $\gamma$, the zero-eigenvalues in association
with the translation of the bounce should be eliminated from the
functional determinant.  For this purpose, we add a small constant $w$
to the function $W(r)$ in Eq.~\eqref{EoM_u} without changing the
bounce. With $w$ being small enough, the functional determinant given
in Eq.~\eqref{Det_J} is proportional to $w^{n_0/2}$, where $n_0$ is
the number of zero-modes.  The zero-eigenvalues can be omitted with
dividing the functional determinant (for non-vanishing $w$) by
$w^{n_0/2}$ and taking $w\rightarrow 0$.

Due to the zero-modes associated with the translation of the bounce,
$A_\phi$ given by Eq.~\eqref{A_varphi} is proportional to $w^{-2}$
(if there is no other zero-mode).  Thus, $A_\phi$ diverges
as $w\rightarrow 0$; such a behavior is related to the infinite
space-time volume.  The dependence of $A_\phi \propto w^{-2}$ originates from
the relation of ${\rm det} [ \phi_{1/2} (r;w)/ \bar{\phi}_{1/2}
(r)]_{r=\infty,w\rightarrow 0}\propto w$, where $\phi_J(r;w)$ obeys
\begin{align}
  \left[ 
    \partial_r^2 + \frac{3}{r} \partial_r 
    - \frac{2J(2J+2)}{r^2} 
    - W(r) - w
  \right] \phi_J(r;w) 
  = 0.
\end{align}
Notice that the zero-modes are involved in the modes with $J=\frac{1}{2}$.  
After omitting the zero-eigenvalues, we obtain
\begin{align}
  A'_\phi = 
  e^{-s_\phi^{(1)}-s_\phi^{(2)}}
  \left[
    \lim_{w\rightarrow 0}
    {\rm det} \left( 
      \frac{\partial_w \phi_{1/2}(r;w)}{\bar{\phi}_{1/2}}
    \right)
    e^{- \tilde{\phi}_{1/2}}
  \right]^{-2}
  \prod_{J\neq 1/2}
  \left.
    \left[
      {\rm det} ( \phi_J / \bar{\phi}_J )
      e^{- \tilde{\phi}_J}
    \right]^{-(2J+1)^2/2}
  \right|_{r=\infty}.
  \label{A'_phi}
\end{align}

Combining Eqs.~\eqref{PrefactorA}, \eqref{A_varphi}, and
\eqref{A'_phi}, the decay rate $\gamma$ is obtained.  Defining
\begin{align}
  S_{\rm tot} \equiv {\cal B} + \Delta S_\phi + \Delta S_\psi,
  \label{Stot}
\end{align}
with 
\begin{align}
  \Delta S_\phi &\equiv 
  - \ln \left[ \frac{{\cal B}^2}{4\pi^2} \frac{A'_\phi}{\Lambda^4} \right],
  \\
  \Delta S_\psi &\equiv
  - \ln A_\psi,  
\end{align}
the decay rate is given by
\begin{align}
  \gamma = \Lambda^4 e^{-S_{\rm tot}},
  \label{eq:decay_rate}
\end{align}
where $\Lambda$ is an arbitrary scale.  Notice that the mass dimension 
in the bracket of $\Delta S_\phi$ is zero, while $\gamma$ is
independent of $\Lambda$.  Taking $\Lambda=100\,{\rm GeV}$, for
example, $S_{\rm tot}$ is required to be larger than $4.0\times 10^2$ to
make the quantity $H_0^{-4}\gamma$ smaller than $1$, where $H_0\simeq
67\,{\rm km/sec/Mpc}$ \cite{Ade:2015xua} is the expansion rate of the
present universe.

The prefactor ${\cal A}$ depends on the renormalization scale via
$e^{-s_\varphi^{(1)}-s_\varphi^{(2)}}$. Such a scale dependence is
necessary to make the decay rate scale-independent.  Indeed, in the
calculation of the decay rate $\gamma={\cal A}e^{-{\cal B}}$, the
renormalization-scale dependence of ${\cal B}$ is compensated by that
of ${\cal A}$.  The cancellation of the scale dependence at the
leading order of $\ln Q$ is shown in the Appendix.\footnote
{Discussion on the scale dependence of the decay rate based on the
effective potential is given in \cite{Camargo-Molina:2013sta}.}
We also comment that $\sum_{p\geq 3}s_\varphi^{(p)}$ can be as large
as $s_\varphi^{(1)}$ and $s_\varphi^{(2)}$.  Thus, the calculation of
both the divergent and convergent parts of the prefactor ${\cal A}$ is
needed.  Explict calculations of the prefactor ${\cal A}$, including
the finite part, are performed in the next sections.

Before closing this section, we comment on the zero-modes in
association with the spontaneous breaking of global symmetry.  In the
following analysis, we also consider the case where a U(1) global
symmetry preserved in the false vacuum is broken by the bounce solution.  
In such a case, another zero-mode appears, which is
related to the U(1) transformation of the bounce.  The path integral
for such a zero-mode can be performed as an integration over the
parameter space of the U(1) group.  The zero-mode is involved in the $J=0$ mode, and its effect
can be taken care of with the following replacement \cite{Kusenko:1996bv}:
\begin{align}
  {\rm det} 
  \left. (\phi_0 / {\bar{\phi}_0})^{-1/2} \right|_{r=\infty}
  \rightarrow
  2 \pi
  \sqrt{\int \frac{ d^4 x }{ 2 \pi } \sum_i q_i^2 \sigma_i^2}
  \left.
    \left[ 
      \lim_{w\rightarrow 0}
      {\rm det} \left( 
        \frac{\partial_w \phi_{0}(r;w)}{\bar{\phi}_{0}(r)}
      \right)
    \right]^{-1/2}
  \right|_{r=\infty},
  \label{Det(GlobalU1)}
\end{align}
where $q_i$ is the charge of the complex scalar field whose real
component contains $\sigma_i$.  The normalization of the U(1)
charge is fixed so that the volume of the U(1) group is equal to
$2\pi$.

\section{Model 1: Model with a Real Scalar Field}
\label{sec:scalar}
\setcounter{equation}{0}

First let us consider the simplest example with a real scalar field
$\Phi=\sigma+\phi$, where $\sigma$ and $\phi$ are the bounce and the
fluctuation around the bounce, respectively. The scalar potential is
\begin{align}
 V (\Phi) =
  -\xi_{\Phi} \Phi +
  \frac{1}{2} m_\Phi^2 \Phi^2 - \frac{1}{2} T_\Phi \Phi^3 + 
  \frac{1}{8} \lambda_\Phi \Phi^4,
  \label{V(Phi)}
\end{align}
with $m_\Phi^2>0$ and $\lambda_\Phi>0$.  The bounce $\sigma$ obeys the
following equation:
\begin{align}
  \partial_r^2 \sigma + \frac{3}{r} \partial_r \sigma - 
  \frac{\partial V(\sigma)}{\partial \sigma} = 0.  
\end{align}
We concentrate on the case where the false vacuum is the one around $\Phi=0$; such a situation is realized for
$T_\Phi^2\gtrsim\lambda_{\Phi} m_{\Phi}^2$.  The renormalization group equations (RGEs) of
the Lagrangian parameters are given by
 \begin{align}
   \frac{d \xi_\Phi}{d \ln Q} &= \frac{3}{16\pi^2}T_\Phi m_\Phi^2,
  \label{dot(xiPhi)}
  \\
  \frac{d m_\Phi^2}{d \ln Q} &= \frac{3}{16\pi^2}(\lambda_\Phi m_\Phi^2+3T_\Phi^2),
  \label{dot(m^2Phi)}
  \\
  \frac{d T_\Phi}{d \ln Q} &= \frac{9}{16\pi^2}\lambda_\Phi T_\Phi,
  \label{dot(APhi)}
  \\
  \frac{d \lambda_\Phi}{d \ln Q} &= \frac{9}{16\pi^2}\lambda_\Phi^2.
  \label{dot(lambdaPhi)}
 \end{align}

We calculate the bounce solution with the potential given
in Eq.~\eqref{V(Phi)} by varying the renormalization scale.  Here, we
adopt the following renormalization condition:
 \begin{align}
  \xi_\Phi (Q_0) &= 0,
  \\
  m_\Phi^2 (Q_0) &= m^2,
  \\
  T_\Phi (Q_0) &= T,
  \\
  \lambda_\Phi (Q_0) &= \lambda,
 \end{align}
with $Q_0 = m$. The parameters $m$ and $\lambda (<1)$ are positive.  
The Lagrangian parameters for different scale are evaluated by using 
the RGEs given in Eqs.~\eqref{dot(xiPhi)}--\eqref{dot(lambdaPhi)}.  
We set $T=m$ and $\lambda=0.6$ in our numerical calculation. 

The model involves various mass scales; the scalar mass is $m$, the 
true vacuum is at $\sigma \simeq 4.2 m$, the potential energy at 
the true vacuum is $|V|^{1/4} \simeq 1.5 m$, the field value at $r=0$ is 
$\sigma(0) \simeq 3.7 m$, and the barrier hight is $|V|^{1/4} \simeq 0.6 m$
at $\Phi \simeq 0.8 m$. 
Since the scales distribute in a wide range, it is difficult 
to determine which is appropriate for the renormalization scale. 
We vary the renormalization scale in the range, $Q/m = 0.5$--5. 
 
\begin{figure}[t]
\begin{center}
\includegraphics[scale=1]{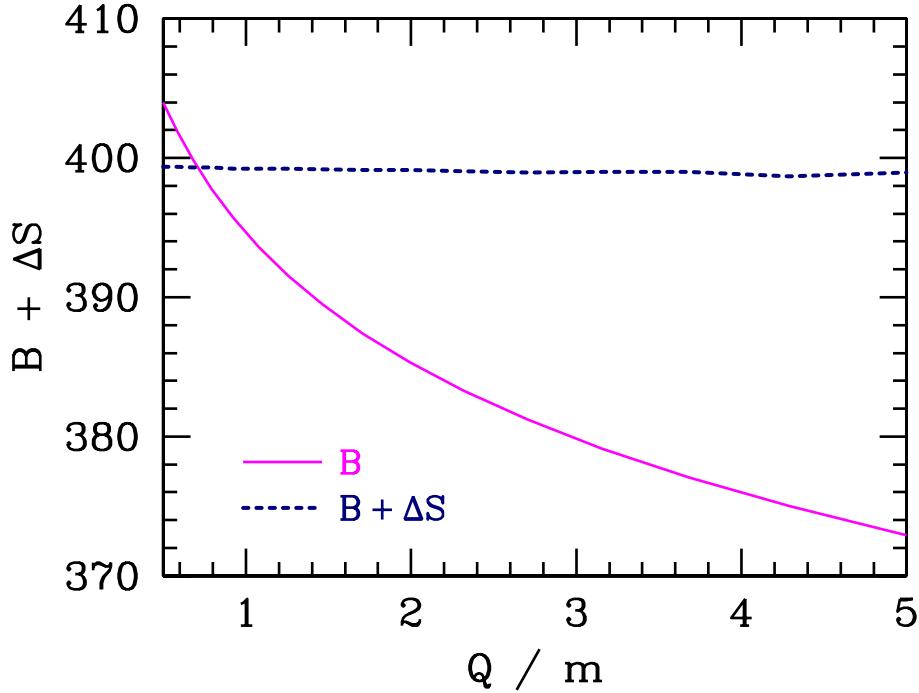}
\end{center}
\caption{${\cal B}$ and $S_{\rm tot}$ as a function of the 
renormalization scale $Q$ in the model of a real scalar field with the
 potential \eqref{V(Phi)}. We take $T=m$ and $\lambda=0.6$.
Also, $\Lambda=100\,{\rm GeV}$ for $\Delta S$.}
\label{fig:toy}
\end{figure}

In Fig.~\ref{fig:toy}, we plot ${\cal B}$ as a function of the 
renormalization scale, $Q$. 
The value of ${\cal B}$ has sizable dependence on $Q$.  
Hence, it is important to properly calculate the prefactor ${\cal A}$
in order to reduce the renormalization-scale uncertainty
as well as to determine the overall scale of the decay rate.

Following the procedure explained in the previous section, we calculate
$S_{\rm tot}$ in Eq.~\eqref{Stot}.  The result is also plotted in Fig.~\ref{fig:toy}.  
When the prefactor ${\cal A}$ is calculated at the one-loop
level, the renormalization-scale uncertainty is significantly reduced; 
${\cal B}$ changes between 404 and 373 for $Q/m = 0.5$--5, while $S_{\rm tot} 
= {\cal B} + \Delta S_\phi$ is stable at $S_{\rm tot} \simeq 400$. 
Thus, the study of this simple model shows that the proper 
inclusion of the prefactor ${\cal A}$ is necessary for an accurate 
estimation of the decay rate $\gamma$.

\section{Model 2: Higgs-Stau System in the MSSM}
\label{sec:mssm}
\setcounter{equation}{0}

In SUSY models, the EWSB vacuum becomes a false 
vacuum if there exists a true vacuum which is CCB or
unbounded-from-below directions.  
The stability of EWSB vacuum often gives significant
constraints on the SUSY parameters \cite{Frere:1983ag,Gunion:1987qv,Casas:1995pd,Kusenko:1996jn,Camargo-Molina:2013sta,Chowdhury:2013dka,Blinov:2013fta,Camargo-Molina:2014pwa,Endo:2015oia}. 
The CCB vacua show up in particular when scalar tri-linear coupling constants are large.
Although the decay rate of the EWSB vacuum is important, 
the prefactor ${\cal A}$ is estimated by an 
order-of-magnitude estimate argument, and is often chosen to be the SUSY scale. 

In this section, we consider the case where the tri-linear coupling of 
the Higgs boson and the scalar taus (staus) is large. 
Such a setup is attractive because, if we assume the universality of the slepton 
masses, SUSY contributions to the muon $g-2$ can be large \cite{Lopez:1993vi,Chattopadhyay:1995ae,Moroi:1995yh}.  
Then, a CCB vacuum may show up in the parameter regions where the muon $g-2$ 
anomaly is solved \cite{Endo:2013lva}.  
We study the decay rate of the EWSB vacuum in such a case.

For simplicity, we consider the case where masses of all the
superparticles and heavy Higgs bosons except for sleptons are much 
larger than the electroweak scale.  We call the mass scale of heavy
superparticles as the SUSY scale $M_{\rm SUSY}$.  Then, an effective 
theory is defined between the electroweak scale and the SUSY scale.  
The effective Lagrangian is described as
\begin{align}
  {\cal L}_{\rm eff} =\ &
  {\cal L}_{\rm kin} 
  - y_t (H q_L t_R^c + {\rm h.c.})
  - m_H^2 |H|^2 
  - \frac{1}{4} \lambda_H |H|^4
  \nonumber \\ &
  - m^2_{\tilde{\ell}_L} |\tilde{\ell}_L|^2
  - m^2_{\tilde{\tau}_R} |\tilde{\tau}_R|^2
  - T_\tau (H^{\dagger} \tilde{\ell}_L \tilde{\tau}_R^* + {\rm h.c.})
  - \frac{1}{4} \kappa^{(1)} |\tilde{\ell}_L|^4
  - \frac{1}{4} \kappa^{(2)} |\tilde{\tau}_R|^4
  \nonumber \\ &
  - \frac{1}{4} \lambda^{(1)} |H|^2 |\tilde{\ell}_L|^2
  - \frac{1}{4} \lambda^{(2)} |H^{\dagger}\tilde{\ell}_L|^2
  - \frac{1}{4} \lambda^{(3)} |H|^2 |\tilde{\tau}_R|^2
  - \frac{1}{4} \kappa^{(3)} |\tilde{\ell}_L|^2 |\tilde{\tau}_R|^2,
  \label{Leff}
\end{align}
where $H$ is the SM-like Higgs doublet, $q_L$ and $t_R^c$ are the
third-generation quark doublet and right-handed anti-top,
respectively, $\tilde{\ell}_L$ is the third-generation slepton doublet,
and $\tilde{\tau}_R$ is the right-handed stau.  We denote the
kinetic terms as ${\cal L}_{\rm kin}$.  Terms containing the
first- and second-generation sleptons are omitted for simplicity
because they are irrelevant for the following discussion.

The scalar potential is significantly affected by the large top-quark 
Yukawa coupling constant $y_t$ and the tri-linear coupling constant 
of the stau $T_\tau$. Because the renormalization-scale dependence of 
$\cal B$ comes from that of the scalar potential, we concentrate on 
the RG evolutions of the couplings associated with the bounce fields.
The relevant RGEs are given by
\begin{align}
  \frac{d m_H^2}{d \ln Q} & = 
  \frac{3y_t^2}{8\pi^2}m_H^2+\frac{1}{8\pi^2}T_{\tau}^2,
  \\
  \frac{d m^2_{\tilde{\ell}_L}}{d \ln Q} & = \frac{1}{8\pi^2} T_{\tau}^2,
  \\
  \frac{d m^2_{\tilde{\tau}_R}}{d \ln Q} & = \frac{1}{4\pi^2}T_{\tau}^2,
  \\
  \frac{d \lambda_H}{d \ln Q} & = 
  \frac{3y_t^2}{4\pi^2} \lambda_H - \frac{3}{8\pi^2}y_t^4,
  \\
  \frac{d T_\tau}{d \ln Q} & = \frac{3y_t^2}{16\pi^2}T_{\tau},
  \\
  \frac{d \lambda^{(I)}}{d \ln Q} & =
  \frac{3y_t^2}{8\pi^2}\lambda^{(I)},
 \\
  \frac{d \kappa^{(I)}}{d \ln Q} & = 0,
\end{align}
with $I=1,2,3$. Because we discuss the renormalization-scale uncertainty 
at the one-loop level, it is sufficient to consider the leading-logarithmic 
dependence on the renormalization scale of the parameters which 
determine the bounce. Hence, we neglect higher loop effects on 
the vacuum decay rate. In particular, the RG running of $y_t$ is neglected 
because the top quark does not compose ${\cal B}$, and thus, the 
RG running is two-loop effects.

In the effective Lagrangian, the parameters associated with the SM  
are determined by the electroweak-scale observables. 
At the top-quark mass scale, we set them as
\begin{align}
  & y_t = \frac{M_t}{v},
  \\
  & m_H^2 (M_t) = - \frac{1}{2} M_h^2,
  \\
  & \lambda_H (M_h) = \frac{M_h^2}{2v^2},
  \label{lambda_H(M_h)}
\end{align}
where $M_t$ and $M_h$ are the top-quark and Higgs masses, respectively. 
Numerically, we use $v\simeq 174\,{\rm GeV}$, $M_t=173.5\,{\rm GeV}$, and
$M_h= 125\,{\rm GeV}$ \cite{Agashe:2014kda}.  With the boundary condition, 
Eq.\,\eqref{lambda_H(M_h)}, $\lambda_H (M_{\rm SUSY})$ may be different 
from the MSSM prediction at the tree level. We assume that such a deviation 
is explained by the threshold correction of the scalar-top loops 
\cite{Okada:1990gg}. 

The quartic scalar coupling constants, $\lambda^{(I)}$ and $\kappa^{(I)}$,
are described by the gauge and tau Yukawa coupling constants at the SUSY scale.  
At the tree level, they are given by
\begin{align}
  \lambda^{(1)}(M_{\rm SUSY}) &= (g^2+g'^2)\cos2\beta,\\
  \lambda^{(2)}(M_{\rm SUSY}) &= 4y_{\tau}^2-2g^2\cos2\beta, \\
  \lambda^{(3)}(M_{\rm SUSY}) &= 4y_{\tau}^2-2g'^2\cos2\beta, \\
  \kappa^{(1)}(M_{\rm SUSY}) &= \frac{1}{2}(g^2+g'^2),\\
  \kappa^{(2)}(M_{\rm SUSY}) &=-\kappa^{(3)}(M_{\rm SUSY}) = 2g'^2,
\end{align}
where $g$ and $g'$ are the gauge coupling constants of the SU(2)$_{\rm L}$ 
and U(1)$_{\rm Y}$ gauge symmetries, respectively, and $\tan\beta$ is 
a ratio of the Higgs vacuum expectation values at the EWSB vacuum, 
$\tan\beta=\langle H_u \rangle/\langle H_d \rangle$. 
In addition, $y_\tau$ is the Yukawa coupling constant of $\tau$ lepton, and is
given by $y_\tau=M_\tau/v$ (with $M_\tau$ being the mass of $\tau$).
The SUSY scale, $M_{\rm SUSY}$, is assumed to be 10\,TeV, and $\tan\beta=20$, 
for our numerical study. 

The stau parameters, $m_{\tilde{\ell}_L}$, 
$m_{\tilde{\tau}_R}$ and $T_\tau$, have not been determined experimentally. 
As one can expect from the Lagrangian Eq.~\eqref{Leff}, CCB
vacua show up when the tri-linear scalar coupling $T_\tau$ becomes
large.  As a sample point at which the EWSB vacuum becomes a false
vacuum, we choose the following parameters,
\begin{align}
  & 
  m_{\tilde{\tau}}\equiv m_{\tilde{\ell}_L} = m_{\tilde{\tau}_R}
  = 250\,{\rm GeV},
  \\ &
  T_\tau = 300\,{\rm GeV}.
\end{align}
at the scale, $Q=m_{\tilde{\tau}}$. Then, the CCB vacuum is
at $\langle H^0\rangle\simeq 1.7\,{\rm TeV}$, 
$\langle\tilde{\tau}_L\rangle\simeq 2.5\,{\rm TeV}$, and
$\langle\tilde{\tau}_R\rangle\simeq 2.5\,{\rm TeV}$, where the
vacuum energy is smaller than that of the EWSB vacuum.

\begin{figure}[t]
\begin{center}
\includegraphics[scale=1]{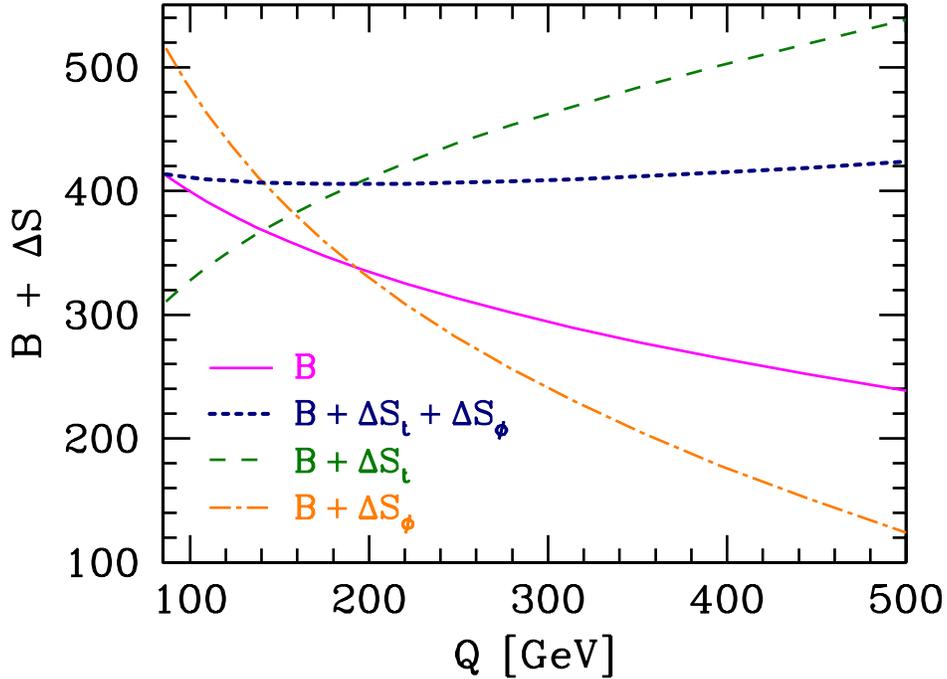}
\end{center}
\caption{Renormalization-scale dependences of ${\cal B}$, 
${\cal B}+\Delta S_t$, ${\cal B}+\Delta S_\phi$, and $S_{\rm tot}
={\cal B}+\Delta S_t+\Delta S_\phi$ in the Higgs-stau model.
Here, $m_{\tilde{\tau}} \equiv m_{\tilde{\ell}_L} = m_{\tilde{\tau}_R}
= 250\,{\rm GeV}, T_\tau = 300\,{\rm GeV}$, and $\tan\beta=20$. 
Also, $\Lambda=100\,{\rm GeV}$ is taken for $\Delta S_\phi$.}
\label{fig:stau}
\end{figure}

In order to see the dependence of ${\cal B}$ on the renormalization 
scale, $Q$ is varied from $M_t/2$ to $2m_{\tilde{\tau}}$.
Using the Lagrangian parameters at the scale $Q$, 
the Euclidean equation of motion is solved to calculate the bounce 
action. In Fig.~\ref{fig:stau}, ${\cal B}$ is plotted as a function 
of $Q$. It changes from 420 to 240 for $Q=M_t/2$ to $2m_{\tilde{\tau}}$, 
corresponding to 45\% scale uncertainty for ${\cal B}=400$. 

The prefactor ${\cal A}$ is calculated by the procedure explained 
in Section \ref{sec:formalism}.  
It consists of the fermion and scalar contributions which are denoted 
as $\Delta S_t$ and $\Delta S_\phi$, respectively;
$\Delta S_t$ comes from the top quark, while $\Delta S_\phi$ is from 
$H$, $\tilde{\ell}_L$, and $\tilde{\tau}_R$.
In our analysis, we neglect the SU(2)$_{\rm L}$ $\times$ U(1)$_{\rm Y}$ 
gauge interactions in the calculation of 
${\cal A}$ because the gauge coupling constants are numerically small.
The inclusion of the gauge boson loops is technically and conceptually complicated,
and is beyond the scope of this paper; this issue will be discussed
elsewhere \cite{EMNS_WorkInProgress}. 
One subtlety is that there exists the U(1)$_{\rm em}$ symmetry which is
preserved in the EWSB vacuum and is broken in the true vacuum.
Because we neglect the U(1)$_{\rm em}$ gauge interaction, the U(1)$_{\rm em}$ 
symmetry is treated as a global symmetry, and
Eq.~\eqref{Det(GlobalU1)} is used to take account 
of the effect of the associated zero-mode.

In Fig.~\ref{fig:stau}, the renormalization-scale dependences of 
${\cal B}+\Delta S_t$, ${\cal B}+\Delta S_\phi$, and $S_{\rm tot}
={\cal B}+\Delta S_t+\Delta S_\phi$ are displayed. 
$\Delta S_t$ and $\Delta S_\phi$ as well as ${\cal B}$
depend on $Q$, and $\Delta S_t$ ($\Delta S_\phi$) increases 
(decreases) as $Q$ increases.
Importantly, the renormalization-scale dependence of $S_{\rm tot}$ is 
significantly reduced. We can see that $S_{\rm tot}$ is stable around 400; 
the scale uncertainty becomes about 5\%.\footnote{
The renormalization-scale dependence can be improved if we take 
all the interactions of the effective Lagrangian such as $\lambda_H$ 
into account for the beta functions.
}
Thus, the proper inclusion of the prefactor ${\cal A}$ stabilizes the decay rate 
of the EWSB vacuum against the change of the renormalization scale.

As in the case of the previous section, 
the calculation of the prefactor ${\cal A}$ is found to be 
important to determine the overall scale of the decay rate
as well as to reduce the renormalization-scale uncertainty.
We also comment that, at the tree level, it is impossible to find an
appropriate renormalization scale to estimate the decay rate of the 
false vacuum, because there is no well-defined procedure to determine 
${\cal A}$ without performing the loop calculation. 

Before closing this section, we comment on other CCB vacua in the MSSM.
They also arise in the stop-Higgs potential \cite{Kusenko:1996jn,Camargo-Molina:2013sta,Chowdhury:2013dka,Blinov:2013fta,Camargo-Molina:2014pwa,Endo:2015oia}.
The calculation of the prefactor ${\cal A}$ in this system has not been performed yet despite of its importance. 
This issue will be discussed elsewhere \cite{EMNS_WorkInProgress}.

\section{Summary}
\label{sec:summary}
\setcounter{equation}{0}

We have performed a detailed calculation of the decay rate of the
false vacuum $\gamma={\cal A} e^{-{\cal B}}$, paying particular
attention to its renormalization-scale dependence.  The bounce action
${\cal B}$ depends on the renormalization scale through the Lagrangian 
parameters, which makes it difficult to accurately calculate the 
decay rate at the tree level.
Such a scale dependence disappears once we take account of the
effects of fluctuations around the bounce, i.e., loop corrections.
In addition, the prefactor ${\cal A}$ cannot be determined 
at the tree level and is often replaced by fourth power of a typical 
mass scale in the Lagrangian.
To resolve this arbitrariness, the calculation of ${\cal A}$ is necessary. 

We have carefully included one-loop corrections to the decay rate.  
We have considered a simple model with a scalar field as well as a
supersymmetric model in which Higgs-stau system has CCB vacua.  With
the change of the renormalization scale within the reasonable range,
the bounce action can change by $O(10)\,\%$ in these models.  
We have shown that the renormalization-scale uncertainty is reduced to 
be $O(1)\,\%$ if the prefactor ${\cal A}$ is taken 
into account properly. Thus, 
for an accurate calculation of the decay rate, proper inclusion of the
loop effects is important.

\

\noindent {\it Acknowledgment}: Y.S. is supported by Grant-in-Aid for
JSPS Fellows under the program number 26-3171.  This work is also
supported by Grant-in-Aid for Scientific research Nos.\ 25105011
(M.E.), 23104008 (T.M.), 26400239 (T.M.), 23104006 (M.M.N.), and
26287039 (M.M.N.), and also by World Premier International Research
Center Initiative (WPI Initiative), MEXT, Japan.

\appendix

\section{Scale Independence at the Leading Order of $\ln Q$}

In this Appendix, we show the cancellation of the $Q$-dependence of
${\cal A}e^{-{\cal B}}$ at the leading order of $\ln Q$.  For
simplicity, we assume that there is no kinetic mixing among scalar
fields, which is the case in the models we have studied.

Let us denote the tree-level potential of real scalar fields as
\begin{align}
  V = \sum_{n=1}^4 \sum_{i_1,\cdots,i_n} C^{(n)}_{i_1,\cdots,i_n} 
  \Phi_{i_1} \cdots \Phi_{i_n},
\end{align}
where $C^{(n)}_{i_1,\cdots,i_n}$ are Lagrangian parameters.  Because
of the renormalizability, terms with $n>4$ do not exist.  The bounce
action is given by
\begin{align}
  {\cal B} = \int d^4 x 
  \left[ 
    - \frac{1}{2} \sum_i \Phi_i \partial^2 \Phi_i
    + V
  \right]_{\Phi\rightarrow\sigma}.
\end{align}

The prefactor ${\cal A}$ depends on $Q$ through $s^{(1)}_\varphi$ and
$s^{(2)}_\varphi$.  The $Q$-dependent parts of $s^{(1)}_\varphi$ and
$s^{(2)}_\varphi$ can be expressed as space-time integrals of local
terms; their sum should result in the following form:
\begin{align}
  \sum_\varphi (s^{(1)}_\varphi + s^{(2)}_\varphi) = 
  \int d^4 x
  \left[ 
    - \frac{1}{2} \sum_i \zeta_i
    \sigma_i \partial^2 \sigma_i
    + \sum_{n=1}^4 \sum_{i_1,\cdots,i_n} \eta^{(n)}_{i_1,\cdots,i_n} 
    \sigma_{i_1} \cdots \sigma_{i_n}
  \right] \ln Q + \cdots.
  \label{s1+s2}
\end{align}
The renormalization-group equations are expressed by using the
wave-function corrections $\zeta_i$ and the vertex corrections
$\eta^{(n)}_{i_1,\cdots,i_n}$.  At the leading order of $\ln Q$, the
scale dependence of the Lagrangian parameters are given by
\begin{align}
  C^{(n)}_{i_1,\cdots,i_n} (Q) = 
  C^{(n)}_{i_1,\cdots,i_n} (Q_0) 
  + \left[
    - \eta^{(n)}_{i_1,\cdots,i_n} 
    + \frac{1}{2} C^{(n)}_{i_1,\cdots,i_n} \sum_{j=i_1}^{i_n} \zeta_j
  \right] \ln (Q/Q_0) + \cdots.
\end{align}

Then, the scale dependence of $S_{\rm tot} = {\cal
  B}+\sum_\varphi\sum_p s^{(p)}_\varphi$ is given by
\begin{align}
  S_{\rm tot} (Q) = S_{\rm tot} (Q_0)
  + \frac{1}{2} \int d^4 x 
  \sum_i \zeta_i 
  \left[
    \Phi_i
    \left(
      - \partial^2 \Phi_i
      + \frac{\partial V}{\partial \Phi_i}
    \right)
  \right]_{\Phi\rightarrow\sigma} \ln (Q/Q_0) + \cdots.
\end{align}
The $Q$-dependent terms in the right-hand side vanish due to the
equation of motion for the bounce.  Notice that, in order to eliminate the
$Q$-dependence from the decay rate, the wave-function corrections as
well as the vertex corrections should be included.

So far, we have neglected the fact that the shape of the bounce
solution $\sigma (r)$ depends on the choice of the renormalization
scale because of the running of the Lagrangian parameters.  Because
the bounce is the solution of the equation of motion, the effect of
the change of the bounce shape on the bounce action ${\cal B}$ is
higher order in $\ln Q$.

\end{document}